\documentclass[journal]{IEEEtran}
\ifCLASSINFOpdf
\else
\fi
\usepackage{multirow}
\usepackage{orcidlink}
\usepackage{adjustbox}
\usepackage{amssymb}
\usepackage{amsmath}
\usepackage{booktabs}
\begin{document}
%
\title{Understanding What Is Not Said:\\Referring Remote Sensing Image Segmentation with Scarce Expressions}
%
%
%

\author{Kai Ye \orcidlink{0000-0003-0914-0213}, Bowen Liu \orcidlink{0009-0000-5006-8925}, Jianghang Lin \orcidlink{0009-0009-3873-3860}, Jiayi Ji \orcidlink{0000-0002-9956-6308}, Pingyang Dai \orcidlink{0000-0001-9780-271X}, Liujuan Cao \orcidlink{0000-0002-7645-9606}
\thanks{ 
Kai Ye, Jianghang Lin, Bowen Liu are with the Key Laboratory of Multimedia Trusted Perception
 and Efficient Computing, Ministry of Education of China, and the Institute
 of Artificial Intelligence, Xiamen University, Xiamen 361005, China (e-mail:
 yekai@stu.xmu.edu.cn; bowenliu@stu.xmu.edu.cn ; hunterjlin007@stu.xmu.edu.cn).

  Jiayi Ji is with the Key Laboratory of Multimedia Trusted Perception and
 Efficient Computing, Ministry of Education of China, Xiamen University,
 Xiamen 361005, China, and also with the National University of Singapore,
 Singapore 119077 (e-mail: jjyxmu@gmail.com).

  Liujuan Cao\textit{(Corresponding author)} and Pingyang Dai are with the Key Laboratory of Multimedia Trusted Perception and Efficient Computing, Ministry of Education
 of China, Xiamen University, Xiamen 361005, China, and also with the
 School of Informatics, Xiamen University, Xiamen 361005, China (e-mail:
 caoliujuan@xmu.edu.cn, pydai@xmu.edu.cn).
 }}

\maketitle

\begin{abstract}
Referring Remote Sensing Image Segmentation (RRSIS) aims to segment instances in remote sensing images according to referring expressions. 
Unlike Referring Image Segmentation on general images, acquiring high-quality referring expressions in the remote sensing domain is particularly challenging due to the prevalence of small, densely distributed objects and complex backgrounds. 
This paper introduces a new learning paradigm, Weakly Referring Expression Learning (WREL) for RRSIS, which leverages abundant class names as weakly referring expressions together with a small set of accurate ones to enable efficient training under limited annotation conditions.  
Furthermore, we provide a theoretical analysis showing that mixed-referring training yields a provable upper bound on the performance gap relative to training with fully annotated referring expressions, thereby establishing the validity of this new setting.  
We also propose LRB-WREL, which integrates a Learnable Reference Bank (LRB) to refine weakly referring expressions through sample-specific prompt embeddings that enrich coarse class-name inputs.  
Combined with a teacher–student optimization framework using dynamically scheduled EMA updates, LRB-WREL stabilizes training and enhances cross-modal generalization under noisy weakly referring supervision.  
Extensive experiments on our newly constructed benchmark with varying weakly referring data ratios validate both the theoretical insights and the practical effectiveness of WREL and LRB-WREL, demonstrating that they can approach or even surpass models trained with fully annotated referring expressions.
\end{abstract}

\begin{IEEEkeywords}
Referring remote sensing image segmentation, referring image segmentation, semi-Supervised Learning, weakly supervised Learning.
\end{IEEEkeywords}

%
\IEEEpeerreviewmaketitle

\begin{figure}
    \centering
    \includegraphics[width=1\linewidth]{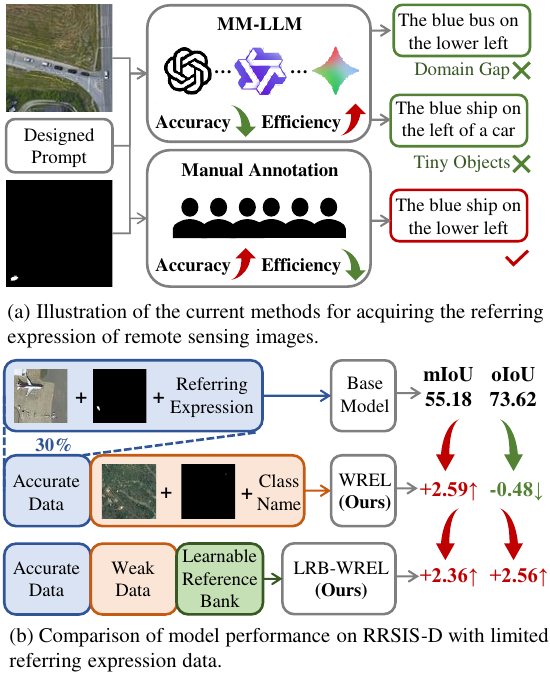}
    \caption{Challenges in obtaining high-quality referring expressions in remote sensing images and the resulting performance gains from supplementing training with class names as weakly referring expressions.}
    \label{fig:placeholder}
\end{figure}

\section{Introduction}
%
%
%
%
\IEEEPARstart{R}{eferring}  Remote Sensing Image Segmentation (RRSIS) has emerged as a growing research focus in recent years~\cite{yuan2024rrsis,lei2024exploring,liu2024rotated,chen2025rsrefseg,liu2025cadformer}.
This task aims to precisely localize and segment target instances in remote sensing images according to given referring expressions, which plays an important role in applications such as land monitoring, disaster assessment, and urban planning.
However, remote sensing images typically contain numerous small-scale and densely distributed objects, making Referring Image Segmentation (RIS) on such data particularly challenging.
Although several datasets and methods have promoted progress in this field~\cite{yuan2024rrsis,liu2024rotated,dong2024cross,yang2025largescalereferringremotesensing,lu2025rrsecs}, the acquisition of high-quality referring expressions remains a major bottleneck.
Existing studies mainly address the lack of mask annotations under weakly supervised or unsupervised settings~\cite{DBLP:journals/tnn/FengZHL24,kim2023shatter,DBLP:conf/cvpr/DaiY24,DBLP:conf/nips/Yang0L0L24,DBLP:conf/icml/YangJM00SJ24}, but systematic research on how to efficiently obtain accurate referring expressions is still lacking.
This problem is even more pronounced in remote sensing scenarios.

As shown in Fig.~\ref{fig:placeholder}(a), referring expressions in remote sensing images are usually obtained in two ways.
One approach leverages multi-modal large language models (e.g., GPT~\cite{radford2018improving,brown2020language}, Qwen~\cite{Qwen2.5-VL}, Gemini~\cite{DBLP:journals/corr/abs-2312-11805}) for automatic generation.
Although efficient, such methods are prone to errors caused by ambiguity of small-objects, complex backgrounds, and domain gaps, thus often requiring human correction.
The other approach relies on manual annotation, which ensures accuracy, but is costly and may introduce subjective bias.
Consequently, high-quality referring expressions are extremely scarce, severely limiting model training and deployment.
Motivated by this, we explore effective training strategies for RRSIS under limited referring expression annotations.

With the rapid progress of large-scale vision models~\cite{kirillov2023segment,radford2021learning,siméoni2025dinov3,liu2023visual,yuan2025sa2vamarryingsam2llava} such as DINO v3~\cite{siméoni2025dinov3} and LLaVA-SAM~\cite{yuan2025sa2vamarryingsam2llava}, it has become increasingly feasible to obtain high-quality masks and instance-level category labels.
This implies that in the image–mask–referring-expression triplet, masks and class names can be acquired at relatively low cost.
Although class names are coarse-grained and may not perfectly align with actual instances, they still provide valuable prior knowledge.
Inspired by weakly supervised and unsupervised learning, we introduce a new learning paradigm termed Weakly Referring Expression Learning for RRSIS (WREL-RRSIS), which leverages class names as weakly referring expressions, combined with a small amount of accurately annotated ones, to enable efficient training under limited referring information.
To this end, we construct a benchmark with varying ratios of weakly referring data, systematically evaluating its impact on model performance, and validating the effectiveness of this new setting.

Furthermore, inspired by the success of learnable prompts in vision–language models~\cite{zhou2022learning,DBLP:conf/cvpr/ZhouYL022,li2024learning,DBLP:conf/cvpr/LiLFZWCY24,DBLP:conf/cvpr/KhattakR0KK23}, we propose LRB-WREL, which incorporates a Learnable Reference Bank (LRB) to compensate for missing fine-grained semantics and imprecise referring cues when using class names as weakly referring expressions.
LRB is introduced after a warm-up stage on accurately annotated data to learn a coarse-to-fine mapping from masks and images to referring expressions.
It provides complementary semantic information for class-name input by learning fine-grained, instance-level localization cues that disambiguate coarse labels.
To mitigate potential bias and overfitting in this mapping, we employ a teacher–student framework, where the student is trained on mixed weak–accurate data, and the teacher—updated through Exponential Moving Average (EMA)~\cite{DBLP:conf/nips/TarvainenV17}—periodically re-optimizes LRB to maintain stability and generalization.
The main contributions of this work are as follows:

\begin{itemize}
    \item We introduce a new learning setting, Weakly Referring Expression Learning for RRSIS (WREL-RRSIS), and validate its effectiveness through both theoretical analysis and experimental validation.
    \item We propose LRB-WREL, which integrates a Learnable Reference Bank and a teacher–student optimization scheme with dynamically scheduled EMA updates to refine weakly referring expressions via sample-specific prompts, stabilizing training and enhancing cross-modal generalization.
    \item We construct a benchmark with varying ratios of weakly and accurately annotated referring data, and demonstrate significant performance improvements across oIoU and mIoU, with results approaching or surpassing training that relies solely on fully annotated referring expressions.
\end{itemize}

\section{Related Work}
\subsection{Referring Remote Sensing Image Segmentation}

Referring Remote Sensing Image Segmentation (RRSIS) aims to segment target objects in remote sensing imagery based on textual referring expressions.  
Existing studies can be broadly categorized into two research lines.  
The first line follows LAVT-based frameworks~\cite{liu2024rotated,lei2024exploring,yuan2024rrsis,shi2025multimodal,liu2025cadformer,ma2025lscf}, among which RMSIN~\cite{liu2024rotated} serves as a representative method featuring multi-scale interaction and fine-grained segmentation.  
FIANet~\cite{lei2024exploring} improves cross-modal alignment by decomposing textual cues into object- and spatial-level representations, while LGCE~\cite{yuan2024rrsis} employs language-guided Transformers to enhance small-object detection.  
MAFN~\cite{shi2025multimodal} introduces cross-modal reasoning for better feature fusion, CADFormer~\cite{liu2025cadformer} explicitly aligns visual and textual features, and LSCF~\cite{ma2025lscf} emphasizes local–global context integration.  
The second line leverages pre-trained vision–language models~\cite{chen2025rsrefseg,chen2025rsrefseg2decouplingreferring}, such as RSRefSeg~\cite{chen2025rsrefseg} combining CLIP and SAM, and RSRefSeg 2~\cite{chen2025rsrefseg2decouplingreferring} adopting a two-stage CLIP–SAM pipeline.  
Although these methods perform well given sufficient annotations, they remain underexplored in remote sensing scenarios where referring expressions are limited and noisy.

\subsection{Weakly- and Semi-supervised Referring Image Segmentation}

Weakly-supervised and semi-supervised Referring Image Segmentation (WRIS/SS-RIS) aims to mitigate the annotation burden associated with pixel-level masks.  
Early WRIS approaches relied on bounding boxes~\cite{DBLP:journals/tnn/FengZHL24}, later evolving toward text-only supervision.  
For instance, SAG~\cite{kim2023shatter} models entity-level relations, Lee et al.~\cite{DBLP:conf/iccv/LeeLNYDT23} enhance consistency via Grad-CAM, and TRIS~\cite{DBLP:conf/iccv/LiuLKXZY0L23} learns text–region correspondences through contrastive maps.  
More recently, PPT~\cite{DBLP:conf/cvpr/DaiY24} employs CLIP and SAM to generate denoised point-based guidance, and PCNet~\cite{DBLP:conf/nips/Yang0L0L24} progressively refines text comprehension for finer alignment.  
In semi-supervised settings, RESMatch~\cite{DBLP:journals/corr/abs-2402-05589} enforces cross-view consistency with pseudo-labels, while SemiRES~\cite{DBLP:conf/icml/YangJM00SJ24} integrates SAM-based confidence-aware refinement.  
Although these methods achieve strong results in natural image domains, their reliance on high-quality textual descriptions and transferable priors limits their applicability to remote sensing imagery, where expressions are sparse, ambiguous, and often refer to numerous small, densely distributed instances.

\subsection{Referring Expression Acquisition}

Acquiring reliable referring expressions remains a critical bottleneck in RRSIS dataset construction.  
RefSegRS~\cite{yuan2024rrsis} first explored template-based automatic generation, which is efficient but produces limited linguistic diversity and unnatural phrasing.  
Subsequent datasets incorporated SAM into the annotation pipeline to improve mask–text alignment:  
RRSIS-D~\cite{liu2024rotated} relies on manually refined SAM masks, RISBench~\cite{dong2024cross} extends SAM with PA-SAM and multi-stage verification, and RefDIOR~\cite{lu2025rrsecs} employs SAM2 with majority-vote fusion to ensure mask–text consistency.  
NWPU-Refer~\cite{yang2025largescalereferringremotesensing} instead uses fully manual annotations for higher naturalness at a significant cost, whereas EarthReason~\cite{li2025segearth} adopts an LLM-based Q\&A strategy still requiring expert correction.  
In summary, automated and semi-automated pipelines improve efficiency but depend on post-correction, while manual annotation guarantees accuracy but remains prohibitively expensive.  
This highlights a central challenge for the field: how to fully exploit existing data and class-level labels to advance RRSIS without relying on additional large-scale manual referring expression construction.
\section{Method}
\subsection{Problem Definition and Theoretical Analysis}

In the Referring Remote Sensing Image Segmentation (RRSIS) task, given a remote sensing image \(I \in \mathbb{R}^{H \times W \times 3}\) and a referring expression \(r\), the goal is to predict the pixel-wise segmentation mask \(M \in \{0,1\}^{H \times W}\) corresponding to the instance indicated by \(r\):
\begin{equation}\label{eq:rrsis}
M = f_\theta(I, r),
\end{equation}
where \(f_\theta\) denotes the segmentation model with parameters \(\theta\).

When high-quality referring expressions are available, each training sample consists of a triplet \((I_i, r_i, M_i)\), where \(r_i\) is an accurate referring expression and \(M_i\) is the corresponding instance mask. However, in large-scale remote sensing applications, collecting high-quality referring expressions is costly and inefficient, since it requires expert annotators to describe numerous small and densely distributed objects and to repeatedly verify their correspondence with instance masks.

To address this issue, we propose Weakly Referring Expression Learning for RRSIS (WREL-RRSIS). In the proposed setting, most image–mask pairs \((I, M)\) contain only class names \(c\) as weakly referring expressions, while only a small subset provides accurate referring expressions. With the advances of large-scale vision models such as DINO v3, which natively supports remote sensing tasks, obtaining instance-level masks and category information from remote sensing images has become highly efficient, providing a solid foundation for the effectiveness and practicality of WREL-RRSIS. We denote the mapping from a class name to a weakly referring expression as
\begin{equation}\label{eq:weak-map}
\tilde{r} = \mathcal{G}(c),
\end{equation}
where \(\mathcal{G}\) is a general transformation that converts the class name \(c\) into a weakly referring expression \(\tilde r\) approximating the accurate referring expression \(r\). The training dataset can thus be expressed as
\begin{equation}\label{eq:data}
\mathcal{D} = \{(I_i, M_i, r_i)\}_{i=1}^{N_a} \cup \{(I_j, M_j, \tilde{r}_j)\}_{j=1}^{N_w},
\end{equation}
where \(N_a\) is the number of samples with accurate referring expressions and \(N_w\) is the number of samples with only weakly referring expressions.

Our objective is to learn parameters \(\theta\) such that the model achieves robust performance under scarce high-quality referring expressions. Accordingly, we formulate a joint loss that combines supervision from both accurate and weakly referring expressions:
\begin{equation}\label{eq:loss}
\mathcal{L} = \frac{1}{N_a}\sum_{i=1}^{N_a}\ell\!\left(f_\theta(I_i, r_i), M_i\right)
+\lambda \frac{1}{N_w}\sum_{j=1}^{N_w}\ell\!\left(f_\theta(I_j, \tilde{r}_j), M_j\right),
\end{equation}
where \(\ell\) denotes the segmentation loss (e.g., cross-entropy), and \(\lambda\) balances the two terms.

\begin{figure*}
    \centering
    \includegraphics[width=1\linewidth]{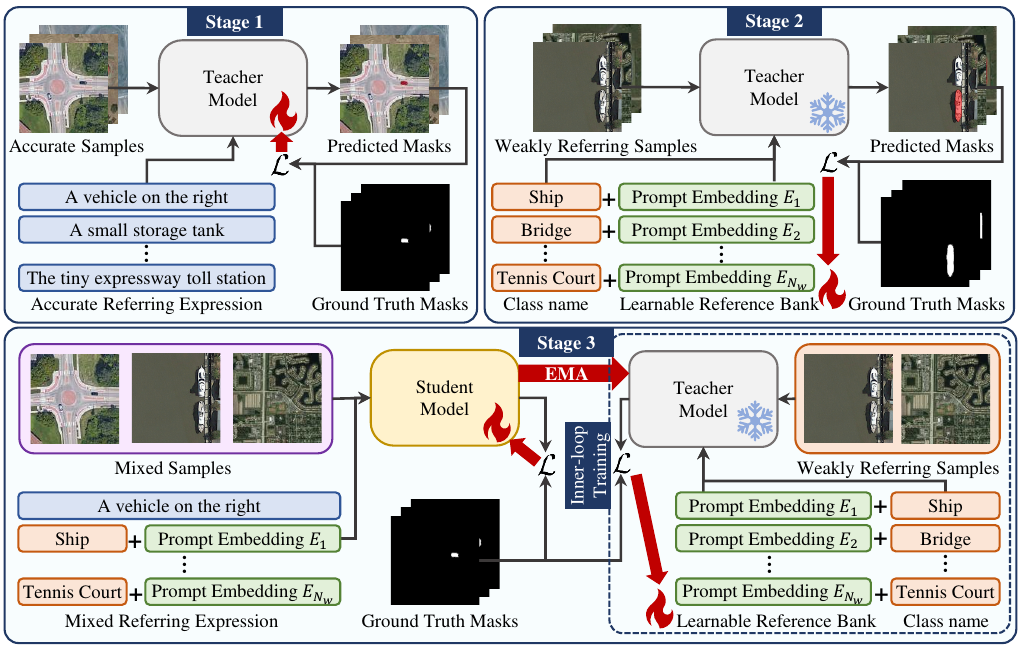}
    \caption{Overview of the proposed LRB-WREL framework, which integrates weakly referring samples with accurate samples through a learnable reference bank and teacher–student optimization.}
    \label{fig:pipeline}
\end{figure*}

Following the analysis of noisy and weak-label learning in \cite{natarajan2013learning,dong2018learning,lu2016learning}, we assume there exists an embedding \(\phi\) such that the approximation error between \(\tilde r\) and the accurate referring expression \(r\) is bounded:
\begin{equation}\label{eq:approx}
\mathbb{E}\!\left[\big\|\phi(\tilde{r})-\phi(r)\big\|^2\right]\le \epsilon.
\end{equation}
Under the Lipschitz continuity of the loss and a bounded hypothesis class, we can bound the gap to the accurate-referring optimum. Let \(\theta^\star\) denote the minimizer of the mixed-referring training objective in Eq.~\eqref{eq:loss} (i.e., trained on a mixture of accurate and weakly referring expressions), and let \(\theta_{a}\) denote the minimizer obtained when all referring expressions are accurate. Specifically, \(\mathcal{R}_{a}(\theta)\) denotes the expected loss of \(f_\theta\) evaluated on the distribution of samples with accurate referring expressions \(p_a\):
\begin{equation}
\mathcal{R}_{a}(\theta)=\mathbb{E}_{(I,r,M)\sim p_{a}}\bigl[\ell(f_\theta(I,r),M)\bigr].
\end{equation}
Then, with probability at least \(1-\delta\),
\begin{equation}\label{eq:bound}
\mathcal{R}_{a}(\theta^\star)-\mathcal{R}_{a}(\theta_{a})
\ \le\ C_1\,\epsilon\ +\ C_2\,\Bigl(\mathfrak{R}_n + \sqrt{\tfrac{\log(1/\delta)}{n}}\Bigr),
\end{equation}
where \(C_1\) depends on the Lipschitz constant of the loss and the continuity of \(\phi\), \(C_2\) depends on the hypothesis class complexity, and \(\mathfrak{R}_n\) denotes the Rademacher complexity. Eq.~\eqref{eq:bound} indicates that as \(\mathcal{G}\) produces weakly referring expressions that better approximate accurate ones (reducing \(\epsilon\)) and as more weakly referring samples become available, the parameters \(\theta^\star\) obtained under mixed-referring training approach the accurate-referring optimum \(\theta_{a}\) in risk measured by \(\mathcal{R}_{a}(\cdot)\). A complete derivation is provided in the Appendix.

WREL-RRSIS explicitly addresses the scarcity of high-quality referring expressions by leveraging abundant masks and class names for efficient training. Mixed-referring training decomposes into an approximation term controlled by \(\epsilon\) and a statistical term controlled by sample size and hypothesis class capacity, jointly ensuring that RRSIS models can achieve performance close to that of training with fully accurate referring annotations even under scarce high-quality referring expressions.

\subsection{Framework Overview}

We propose a training framework named Learnable Reference Bank-based Weakly Referring Expression Learning for RRSIS (LRB-WREL), designed to exploit abundant weakly referring expressions under scarce high-quality annotations, thereby progressively enhancing RRSIS performance.

At the outset, the model is pre-trained on a small set of accurate referring expression samples to acquire an initial language–vision alignment under precise referring annotations, yielding the initial model parameters \(\theta_{0}\). 
Subsequently, a Learnable Reference Bank (LRB) is introduced and optimized on the frozen warm-up model to provide supplementary semantics for weakly referring samples. 
Trainable prompt vectors are injected into reserved slots of the text embedding sequence, and the attention mask is synchronously updated. This transforms weakly referring expressions based solely on class names into more discriminative and contextually aligned representations before being encoded and fused with visual features for segmentation supervision.

Subsequently, a teacher–student architecture is adopted to jointly optimize the model and the reference bank.
The student model is trained on a mixed dataset that integrates LRB-enhanced weakly referring samples with accurate referring samples.
Meanwhile, the teacher model tracks the student parameters via a dynamically scheduled EMA to provide stable supervision.
The reference bank is iteratively updated on the frozen teacher through inner-loop optimization, which further refines weakly referring expressions during training.
This closed-loop process enables a cooperative convergence among the student, the teacher, and the LRB.
It stabilizes training with mixed accurate and weakly referring expressions, improves the model’s ability to map weakly referring expressions to fine-grained semantics, and improves cross-modal generalization.

\subsection{Learnable Reference Bank}

Weakly referring expressions that rely solely on class names often lack the semantic detail and referential specificity required to distinguish fine-grained targets. Moreover, the same class name may correspond to different semantics under varying contexts, such as scale, object density, or viewpoint, leading to unstable referring cues. To address this issue, we introduce a Learnable Reference Bank (LRB), which provides sample-specific semantic augmentation without additional annotation cost. By parameterizing each weakly referring sample with trainable prompt embeddings, the LRB upgrades coarse class-name inputs into more discriminative referring embeddings before entering the text encoder.

Formally, for each weakly referring sample indexed by \(j\in\{1,\dots,N_{w}\}\), the LRB maintains a prompt matrix \(P\in\mathbb{R}^{N_{w}\times p\times d}\), where \(p\) is the number of trainable prompt tokens assigned to the sample and \(d\) is the dimensionality of each token (aligned with the hidden size of the text encoder). Given the text embedding sequence \(X_{j}\in\mathbb{R}^{L\times d}\) and attention mask \(A_{j}\in\{0,1\}^{L}\), we define the padding set as
\begin{equation}
\Omega_j=\{\,l\mid A_j[l]=0\,\}.
\end{equation}
The prompt injection process replaces these reserved slots with \(P_{j}\) while updating the corresponding attention mask:
\begin{equation}
(\bar X_j,\bar A_j)=\mathrm{Fill}(X_j,A_j;P_j),
\end{equation}
where for each token position \(l\),
\begin{equation}
\begin{aligned}
\bar X_j[l]=
\begin{cases}
P_j[l], & l\in\Omega_j,\\
X_j[l], & \text{otherwise},
\end{cases}
\\
\bar A_j[l]=
\begin{cases}
1, & l\in\Omega_j,\\
A_j[l], & \text{otherwise}.
\end{cases}
\end{aligned}
\end{equation}
In this way, the injected representation \((\bar X_j,\bar A_j)\) preserves the original token order and length, while semantically enriching weakly referring expressions at positions that would otherwise carry no information.

The referring embedding enhanced by LRB is then defined as
\begin{equation}
\tilde r_{j}=\mathrm{Encode}(\bar X_{j},\bar A_{j}),
\end{equation}
which can be equivalently written as
\begin{equation}
\mathcal{G}_{\text{LRB}}(X_j,A_j;P_{j})
=\mathrm{Encode}(\mathrm{Fill}(X_j,A_j;P_{j})).
\end{equation}
The resulting \(\tilde r_j\) is fused with image features through the segmentation model:
\begin{equation}
\hat M_{j}=f_\theta(I_{j},\tilde r_{j}).
\end{equation}

This formulation highlights two advantages of the LRB. First, the injection process is minimally invasive, since only padding slots are used, thereby retaining the pre-trained encoder’s positional alignment and token structure. Second, sample-specific prompts \(P_j\) explicitly capture contextual factors such as surrounding clutter, scale, or texture variations, which are especially critical in remote sensing scenarios where objects are small and densely distributed.

During training, to prevent noise in weakly referring expressions from destabilizing the backbone, the encoder and segmentation model are frozen, and only the prompt matrices \(P\) are optimized. This ensures that LRB learns a stable mapping from class names to fine-grained referring embeddings within a fixed semantic space. Empirically, this design allows coarse class-name references to be systematically upgraded into more discriminative embeddings, providing reliable supervision even when accurate referring expressions are scarce.

\subsection{Update Strategy}

During training, the Learnable Reference Bank (LRB), the student network, and the teacher network are updated in a closed-loop manner so that weakly referring samples are progressively refined without destabilizing the main segmentation model.
Specifically, in each mini-batch, let \(B=S\cup W\) be the index set of all samples, where \(S\) denotes the subset with accurate referring expressions and \(W\) the subset with weakly referring expressions.
For weakly referring samples, the LRB parameters \(\{P_{j}\}_{j\in W}\) are optimized by mask-level supervision under a frozen teacher model, which preserves the pre-trained semantic space while allowing the bank to learn a mapping from coarse class names to fine-grained referring embeddings. The objective for this inner-loop calibration is
\begin{equation}
\min_{\{P_{j}:j\in W\}}
\frac{1}{|W|}\sum_{j\in W}
\ell\!\Bigl(
f_{\theta_{T}}\bigl(I_{j},
\mathcal{G}_{\text{LRB}}(X_{j},A_{j};P_{j})\bigr),M_{j}
\Bigr),
\end{equation}
where \(\theta_{T}\) is the frozen teacher parameters.

After calibrating the bank on weakly referring samples, the student network parameters \(\theta_{S}\) are updated on the entire batch by mixing accurate referring expressions with stop-gradient LRB-enhanced weakly referring expressions for weak samples. This ensures that gradients flow only to the student while the bank remains fixed during this stage. The student update is given by
\begin{equation}
\min_{\theta_{S}}\;\frac{1}{|B|}\sum_{i\in B}
\ell\!\Bigl(
f_{\theta_{S}}\bigl(I_{i},\, r^{\mathrm{mix}}_{i}\bigr),\; M_{i}
\Bigr),
\end{equation}
where the mixed referring embedding \(r^{\mathrm{mix}}_{i}\) is defined as
\begin{equation}
r^{\mathrm{mix}}_{i}=
\begin{cases}
\mathrm{sg}\!\bigl[\mathcal{G}_{\text{LRB}}(X_{i},A_{i};P_{i})\bigr], & i\in W,\\[0.5em]
\mathrm{Encode}(X_{i},A_{i}), & i\in S.
\end{cases}
\end{equation}
Here \(\mathrm{sg}[\cdot]\) denotes the stop-gradient operator that prevents gradients from propagating back to the LRB during the student update, so only the student model learns from the enhanced referring embeddings.

Finally, the teacher network parameters are updated as an exponential moving average (EMA) of the student parameters:
\begin{equation}
\theta_{T}\leftarrow \alpha_{t}\,\theta_{T}+(1-\alpha_{t})\,\theta_{S}.
\end{equation}
The momentum coefficient \(\alpha_{t}\) is not fixed but scheduled dynamically as training progresses: it starts small so that the teacher can quickly track improvements in the student early on, and gradually increases to a high value (e.g., 0.9995) to yield a stable teacher later in training. This dynamic EMA produces a smoothed target distribution under noisy weakly referring expressions, stabilizing convergence and ensuring that the student, teacher, and reference bank cooperate effectively throughout training.

\begin{table*}[htbp]
  \centering
  \caption{Quantitative results of LRB-WREL and related settings on RRSIS-D with varying ratios of accurate referring expressions.}
  \label{LRB-rrsis-d}
  \begin{adjustbox}{width=\textwidth}
    \begin{tabular}{c|c|cccccccccc|cccc}
    \toprule
    \multirow{2}[2]{*}{Ratio} & \multirow{2}[2]{*}{Method} & \multicolumn{2}{c}{P@0.5} & \multicolumn{2}{c}{P@0.6} & \multicolumn{2}{c}{P@0.7} & \multicolumn{2}{c}{P@0.8} & \multicolumn{2}{c|}{P@0.9} & \multicolumn{2}{c}{oIOU} & \multicolumn{2}{c}{mIOU} \\
          &       & Val   & Test  & Val   & Test  & Val   & Test  & Val   & Test  & Val   & Test  & Val   & Test  & Val   & Test \\
    \midrule
    100\% & RMSIN & 74.66 & 74.26 & 68.22 & 67.25 & 57.41 & 55.93 & 45.29 & 42.55 & 24.43 & 24.53 & 78.27 & 77.79 & 65.10 & 64.20 \\
    \midrule
    \multirow{3}[2]{*}{10\%} & Only Accurate & 45.80 & 45.76 & 38.16 & 37.66 & 30.29 & 29.59 & 20.34 & 20.63 & 8.45  & 8.62  & 62.56 & 63.03 & 41.34 & 41.15 \\
          & WREL (ours) & 61.38 & 62.42 & 55.06 & 56.56 & \textbf{46.84} & 46.25 & \textbf{36.26} & \textbf{35.59} & \textbf{20.98} & \textbf{20.20} & 69.95 & 71.48 & 54.15 & 54.63 \\
          & LRB-WREL (ours) & \textbf{65.11} & \textbf{63.69} & \textbf{57.13} & \textbf{56.88} & 46.67 & \textbf{46.39} & \textbf{36.26} & 34.79 & 20.11 & 19.76 & \textbf{72.27} & \textbf{71.70} & \textbf{56.87} & \textbf{56.32} \\
    \midrule
    \multirow{3}[2]{*}{30\%} & Only Accurate & 62.76 & 61.19 & 54.48 & 53.78 & 45.06 & 43.75 & 34.54 & 32.81 & 17.13 & 17.81 & 73.62 & 72.34 & 55.18 & 54.07 \\
          & WREL (ours) & 65.92 & \textbf{66.91} & 59.83 & \textbf{60.44} & 48.74 & 49.04 & 37.01 & 37.23 & 20.63 & 20.91 & 73.24 & 74.28 & 57.77 & 58.19 \\
          & LRB-WREL (ours) & \textbf{68.62} & 66.45 & \textbf{61.49} & 59.78 & \textbf{50.63} & \textbf{49.99} & \textbf{39.37} & \textbf{37.69} & \textbf{21.61} & \textbf{21.14} & \textbf{75.80} & \textbf{74.32} & \textbf{60.13} & \textbf{58.79} \\
    \midrule
    \multirow{3}[2]{*}{50\%} & Only Accurate & 67.64 & 63.80 & 60.23 & 55.73 & 49.60 & 44.64 & 37.64 & 32.98 & 21.49 & 16.55 & 75.65 & 73.51 & 59.42 & 56.18 \\
          & WREL (ours) & 70.17 & \textbf{71.33} & 62.87 & \textbf{63.89} & 51.84 & \textbf{52.46} & 39.43 & 39.21 & 22.59 & 22.75 & 76.15 & 76.50 & 61.31 & 61.42 \\
          & LRB-WREL (ours) & \textbf{71.90} & 70.96 & \textbf{63.91} & 63.86 & \textbf{53.22} & 52.26 & \textbf{40.52} & \textbf{40.16} & \textbf{22.93} & \textbf{22.95} & \textbf{76.42} & \textbf{76.57} & \textbf{62.69} & \textbf{62.08} \\
    \bottomrule
    \end{tabular}%
    \end{adjustbox}
\vspace{1mm}
\footnotesize{
“Only Accurate” denotes the RMSIN baseline trained only on accurate expressions. “WREL” uses a combination of accurate and weakly referring expressions, while “LRB-WREL” further introduces a Learnable Reference Bank and teacher–student optimization to enhance weak references.}
\end{table*}%
\begin{table*}[htbp]
  \centering
  \caption{Quantitative results of LRB-WREL and related settings on RIS-LAD with varying ratios of accurate referring expressions.}
  \label{LRB-lad}
  \begin{adjustbox}{width=\textwidth}
    \begin{tabular}{c|c|cccccccccc|cccc}
    \toprule
    \multirow{2}[2]{*}{Ratio} & \multirow{2}[2]{*}{Method} & \multicolumn{2}{c}{P@0.5} & \multicolumn{2}{c}{P@0.6} & \multicolumn{2}{c}{P@0.7} & \multicolumn{2}{c}{P@0.8} & \multicolumn{2}{c|}{P@0.9} & \multicolumn{2}{c}{oIOU} & \multicolumn{2}{c}{mIOU} \\
          &       & Val   & Test  & Val   & Test  & Val   & Test  & Val   & Test  & Val   & Test  & Val   & Test  & Val   & Test \\
    \midrule
    100\% & RMSIN & 45.85 & 43.36 & 40.97 & 38.11 & 35.24 & 32.36 & 28.01 & 25.08 & 16.33 & 13.75 & 50.17 & 48.82 & 42.08 & 39.60 \\
    \midrule
    \multirow{3}[1]{*}{10\%} & Only Accurate & 6.66  & 5.90  & 4.44  & 3.44  & 2.65  & 2.14  & 1.07  & 0.94  & 0.21  & 0.22  & 14.33 & 13.69 & 9.04  & 8.32 \\
          & WREL (ours) & 18.41 & 16.61 & 14.83 & 13.61 & 11.96 & 10.68 & \textbf{9.74} & 7.56  & 3.87  & \textbf{4.02} & 25.74 & 26.10 & 18.88 & 16.90 \\
          & LRB-WREL (ours) & \textbf{20.85} & \textbf{20.27} & \textbf{16.83} & \textbf{16.32} & \textbf{13.97} & \textbf{12.27} & \textbf{9.74} & \textbf{8.98} & \textbf{4.58} & 3.73  & \textbf{27.24} & \textbf{28.13} & \textbf{22.20} & \textbf{21.05} \\
    \midrule
    \multirow{3}[1]{*}{30\%} & Only Accurate & 24.36 & 21.97 & 19.77 & 17.34 & 14.54 & 13.07 & 10.32 & 9.34  & 3.65  & 4.34  & 33.49 & 33.17 & 23.30 & 22.07 \\
          & WREL (ours) & \textbf{30.37} & 26.89 & \textbf{25.93} & \textbf{22.91} & \textbf{20.77} & \textbf{19.11} & \textbf{15.26} & \textbf{13.97} & 6.95  & \textbf{7.13} & 34.76 & 34.71 & 28.49 & 26.04 \\
          & LRB-WREL (ours) & 29.23 & \textbf{27.36} & 24.57 & 22.29 & 18.84 & 17.55 & 14.11 & 12.63 & \textbf{7.02} & 6.37  & \textbf{36.10} & \textbf{35.11} & \textbf{30.20} & \textbf{28.42} \\
    \midrule
    \multirow{3}[2]{*}{50\%} & Only Accurate & 31.66 & 27.29 & 27.22 & 22.58 & 22.21 & 18.53 & 15.26 & 13.36 & 6.73  & 6.26  & 39.08 & 37.34 & 30.53 & 27.35 \\
          & WREL (ours) & 32.74 & 30.8  & 27.87 & 25.15 & 23.57 & 20.52 & 18.19 & 15.31 & 9.03  & 7.96  & 38.43 & 39.14 & 30.54 & 28.68 \\
          & LRB-WREL (ours) & \textbf{34.60 } & \textbf{31.45} & \textbf{30.01} & \textbf{26.20 } & \textbf{25.57} & \textbf{22.44} & \textbf{19.63} & \textbf{16.43} & \textbf{9.81} & \textbf{8.43} & \textbf{40.29} & \textbf{39.59} & \textbf{33.06} & \textbf{30.36} \\
    \bottomrule
    \end{tabular}%
    \end{adjustbox}
\end{table*}%
\section{Experiments}
\subsection{Implementation Details}

\paragraph{Experimental Settings.}
We adopt RMSIN as our base model, primarily because it is a representative work of the LAVT-based architecture in Referring Remote Sensing Image Segmentation (RRSIS). As highlighted in related studies, RMSIN not only pioneered the adaptation of LAVT to the RRSIS task but also inspired many subsequent approaches that followed its architectural design.

In the first stage, the visual backbone is a Swin Transformer~\cite{DBLP:conf/iccv/LiuL00W0LG21} pre-trained on ImageNet-22K~\cite{DBLP:conf/cvpr/DengDSLL009}, while the language backbone employs the BERT-base model from HuggingFace’s library~\cite{DBLP:conf/emnlp/WolfDSCDMCRLFDS20}. The model is trained for 15 epochs using the AdamW optimizer~\cite{loshchilovdecoupled} with a weight decay of 0.01 and an initial learning rate of \(3\times10^{-5}\), scheduled by polynomial decay. Training is conducted on four RTX 3090 GPUs with a batch size of 8.

In the second stage, where the Learnable Reference Bank (LRB) is warmed up, the initial learning rate is reduced to \(1\times10^{-5}\), and training proceeds for 40 epochs to stabilize the learning of prompt embeddings. 
In the third stage, the student and teacher models are trained jointly. The student model is optimized for 40 epochs with a starting learning rate of \(3\times10^{-5}\). The LRB is updated once per training step, with a learning rate of \(1\times10^{-6}\), and each prompt parameter \(P_j\) is updated for one inner step. The teacher model is updated via a dynamic Exponential Moving Average (EMA) of the student parameters, with details of the scheduling strategy provided in the Appendix.

\paragraph{Evaluation Metrics.}
We adopt Overall Intersection-over-Union (oIoU), Mean Intersection-over-Union (mIoU), and Precision@X (P@X) as evaluation metrics, following prior studies. Together, these metrics capture both global accuracy and instance-level precision under different IoU thresholds, providing a comprehensive evaluation of model performance in the presence of mixed accurate and weakly referring expressions.


\paragraph{Benchmark Design.}
To systematically evaluate algorithm performance under the Weakly Referring Expression Learning for RRSIS (WREL-RRSIS) setting, we construct a new benchmark based on two representative RRSIS datasets. 
The first is RRSIS-D, which is one of the most widely used datasets in this field. It provides a comprehensive reflection of the general scenarios and challenges faced by RIS in remote sensing imagery, making it an appropriate baseline for overall evaluation. 
The second is RIS-LAD, a dataset focusing on low-altitude UAV imagery. Due to its highly dense distribution of same-category objects, the quality of referring expressions has a particularly strong impact on model performance, making it a suitable testbed for examining algorithm behavior under the WREL-RRSIS setting.

In building the benchmark, we simulate the scarcity of high-quality referring expressions by creating three ratio configurations of accurate to weakly referring samples: 1:9, 3:7, and 5:5. 
Importantly, the division is performed at the category level to ensure that all classes are represented in the accurate-referring subset, thereby avoiding potential bias caused by missing categories.

\subsection{Experimental Results and Analysis}
Table~\ref{LRB-rrsis-d} presents the quantitative results on the RRSIS-D dataset under different ratios of accurate and weakly referring samples. 
Overall, both WREL and LRB-WREL deliver substantial performance improvements compared with the “Only Accurate” baseline, demonstrating the effectiveness of leveraging weakly referring expressions to compensate for the scarcity of high-quality referring data. 
In particular, when accurate referring expressions are extremely limited (10\%), LRB-WREL achieves increases of 8.67\% in oIoU and 15.17\% in mIoU on the test subset over the “Only Accurate” baseline, and further shows a more pronounced improvement on the validation set, reaching gains of 7.55\% in oIoU and 15.53\% in mIoU compared with WREL.
These results show that weakly referring samples, though potentially noisy, constitute effective complementary signals that improve performance under WREL, while the Learnable Reference Bank further strengthens these benefits by enhancing semantic specificity and restoring fine-grained information.

As the proportion of accurate referring expressions increases from 10\% to 50\%, the performance gains over the “Only Accurate” baseline remain substantial, indicating that weakly referring samples continue to provide complementary supervisory signals rather than acting as mere noise.
Meanwhile, the gap between WREL and LRB-WREL gradually narrows as the annotation ratio grows, suggesting that with sufficient accurate referring expressions the model can already learn strong referential cues, thereby reducing the marginal benefit of LRB’s fine-grained semantic refinement.
This trend supports our design: LRB-WREL is particularly advantageous in the low-annotation regime, while both WREL and LRB-WREL maintain clear advantages as accurate supervision becomes more abundant.

On RIS-LAD, which contains densely distributed same-category objects and thus requires finer referential cues for accurate localization, both WREL and LRB-WREL show clear improvements over the baseline trained only with accurate referring expressions, as presented in Table~\ref{LRB-lad}. 
Under the 10\% accurate setting, where annotation scarcity is most severe, WREL already provides a notable boost, while LRB-WREL further raises test oIoU and mIoU to 28.13\% and 21.05\%, respectively. This demonstrates its stronger capability in recovering fine-grained semantic alignment when detailed linguistic guidance is crucial.  
As the proportion of accurate data increases to 30\% and 50\%, both approaches maintain steady gains over the baseline, with LRB-WREL consistently outperforming WREL. Notably, the performance gap between the two methods becomes more evident on RIS-LAD than on RRSIS-D, indicating that the Learnable Reference Bank is particularly effective in scenes demanding precise referential reasoning.  
These results confirm that leveraging weakly referring expressions provides reliable complementary information, and that the proposed LRB-WREL remains robust and beneficial even as accurate referring annotations become more abundant.

\subsection{Ablation Study}
To investigate the effectiveness of different training strategies, we conduct a series of ablation experiments on the test set.
\paragraph{Inner-loop Training Step.}  
We first investigate the influence of the inner-loop update steps in optimizing the Learnable Reference Bank.  
As reported in Table~\ref{xiaorong1}, disabling LRB updates (\textit{Step=0}) leads to a noticeable performance drop, with the model reaching only 76.29 oIoU and 61.43 mIoU. Introducing a single-step inner update significantly improves performance to 76.42 oIoU and 62.69 mIoU, confirming that moderate LRB optimization effectively enhances weakly referring representations by refining their semantic alignment.  
However, further increasing the number of inner updates to three or five steps yields no additional benefit and even slightly degrades the results. This can be attributed to overfitting of the LRB parameters to local noise from weakly referring expressions, which disrupts the stability of joint optimization with the student model.  

\begin{table}[htbp]
  \centering
  \caption{Ablation study on the effect of internal parameter updates.}
  \label{xiaorong1}
  \resizebox{\linewidth}{!}{%
    \begin{tabular}{c|ccccc|cc}
    \toprule
    Step & P@0.5 & P@0.6 & P@0.7 & P@0.8 & P@0.9 & oIOU  & mIOU \\
    \midrule
    0    & 70.63 & 63.45 & 52.24 & 40.06 & 22.01 & 76.29 & 61.43 \\
    1     & 71.90 & 63.91 & 53.22 & 40.52 & 22.93 & 76.42 & 62.69 \\
    3     & 70.52 & 63.39 & 51.90 & 39.89 & 22.30 & 75.92 & 62.34 \\
    5     & 70.46 & 63.74 & 52.76 & 39.77 & 21.67 & 76.47 & 61.67 \\
    \bottomrule
    \end{tabular}%
    }
\end{table}%
\paragraph{Update Frequency.}
We further investigate the effect of varying the update frequency of internal parameters in the Learnable Reference Bank.  
As summarized in Table~\ref{xiaorong2}, performing updates every batch (Freq=1) yields the highest oIoU (76.42) and mIoU (62.69), while reducing the update frequency to every 3 or 5 batches slightly decreases both metrics.  
This suggests that frequent parameter updates help stabilize cross-modal alignment during training, allowing the model to continuously refine weakly referring representations based on the most recent student feedback.  
In contrast, less frequent updates cause a lag in parameter synchronization, weakening the consistency between textual and visual embeddings and slowing the propagation of semantic cues across modalities.
\begin{table}[htbp]
  \centering
  \caption{Ablation study on the frequency of internal parameter updates.}
  \label{xiaorong2}
  \resizebox{\linewidth}{!}{%
    \begin{tabular}{c|ccccc|cc}
    \toprule
    Freq & P@0.5 & P@0.6 & P@0.7 & P@0.8 & P@0.9 & oIOU  & mIOU \\
    \midrule
    1     & 71.90 & 63.91 & 53.22 & 40.52 & 22.93 & 76.42 & 62.69 \\
    3     & 71.90 & 64.14 & 52.64 & 39.66 & 22.30 & 76.25 & 62.42 \\
    5     & 71.72 & 63.68 & 52.76 & 39.94 & 22.59 & 76.42 & 62.32 \\
    \bottomrule
    \end{tabular}%
    }
\end{table}%
\paragraph{Warm-up Epochs.}  
Finally, we analyze the impact of the warm-up duration used for initializing LRB and student model.  
As shown in Table~\ref{xiaorong3}, using 10 or 20 warm-up epochs produces comparable results to the 15-epoch baseline, with the 10-epoch setting achieving the highest oIoU (76.93\%) but a slightly lower mIoU (62.30\%).  
This indicates that an appropriate warm-up phase allows the model to establish a stable mapping from image and mask features to referring expressions, enabling LRB to learn reliable semantic correlations.  
However, insufficient warm-up may result in an underdeveloped mapping, limiting the model’s ability to align visual and textual cues, while excessive warm-up can cause overfitting to limited supervision, thereby degrading the generalization of LRB.  
These findings suggest that moderate pre-training of internal updates is crucial for stabilizing convergence and ensuring high-quality representation learning.
\begin{table}[htbp]
  \centering
  \caption{Ablation study on the number of warm-up epochs for internal updates.}
  \label{xiaorong3}
  \resizebox{\linewidth}{!}{%
    \begin{tabular}{c|ccccc|cc}
    \toprule
    Epoch & P@0.5 & P@0.6 & P@0.7 & P@0.8 & P@0.9 & oIOU  & mIOU \\
    \midrule
    10    & 71.32 & 63.45 & 53.68 & 39.60 & 22.41 & 76.93 & 62.30 \\
    15    & 71.90 & 63.91 & 53.22 & 40.52 & 22.93 & 76.42 & 62.69 \\
    20    & 70.69 & 62.76 & 52.53 & 40.34 & 22.59 & 76.42 & 62.18 \\
    \bottomrule
    \end{tabular}%
    }
\end{table}%

\section{Conclusion}
This work introduces Weakly Referring Expression Learning (WREL) as a new training setting for Referring Remote Sensing Image Segmentation (RRSIS), addressing the challenge of scarce high-quality referring annotations.  
By leveraging class names as weakly referring expressions, WREL demonstrates that easily obtainable weak data can effectively supplement limited accurate descriptions, providing a practical and scalable solution for large-scale remote sensing scenarios.  
On top of this setting, we further propose LRB-WREL, which integrates a Learnable Reference Bank and a teacher–student update strategy to refine weak expressions into fine-grained, context-aware representations.  
Extensive experiments on two benchmarks, RRSIS-D and RIS-LAD, validate that both WREL and LRB-WREL significantly improve segmentation accuracy, especially in data-scarce and fine-grained referential conditions.  
Overall, this study opens a new perspective for annotation-efficient RRSIS by showing that structured weak supervision can bridge the gap between coarse class-level cues and precise referring understanding.

\ifCLASSOPTIONcaptionsoff
  \newpage
\fi



\bibliographystyle{IEEEtran}
\bibliography{bibtex/bib/IEEEexample}
%


%




\end{document}